\documentclass[review, times]{elsarticle}
\usepackage{url} 
\usepackage{lineno}
\usepackage[inline]{trackchanges} 
\usepackage{amsmath}
\usepackage{commath}
\usepackage{amssymb}
\usepackage{amsbsy}
\usepackage[utf8]{inputenc}
\usepackage{nomencl}
\usepackage{bbold}

\makeatletter
\def\ps@pprintTitle{%
  \let\@oddhead\@empty
  \let\@evenhead\@empty
  \let\@oddfoot\@empty
  \let\@evenfoot\@oddfoot
}
\makeatother

\journal{Advances in Water Resources}
 \usepackage{graphicx,caption}
 \usepackage[export]{adjustbox}
 \usepackage{wrapfig}

\begin{document}



\title{Singular Value Decomposition for Single-Phase Flow and Cluster Identification in Heterogeneous Pore Networks \footnote{Ben-Noah, I., Hidalgo, J. J., \& Dentz, M. (2024). Singular value decomposition for single-phase flow and cluster identification in heterogeneous pore networks. Advances in Water Resources, 192, 104779.}}






\author[inst1,inst2]{Ilan Ben-Noah}
\author[inst1]{Juan J. Hidalgo}
\author[inst1]{Marco Dentz}

\affiliation[inst1]{Institute of Environmental Assessment and Water Research (IDAEA), Spanish National Research Council (CSIC), Barcelona, Spain.}
\affiliation[inst2]{Department of Environmental Physics and Irrigation, Institute of Soil, Water and Environmental Sciences, The Volcani Institute, Agricultural Research Organization, Rishon LeZion, Israel}

\begin{frontmatter}

\begin{abstract}
Pore networks play a key role in understanding and quantifying flow and transport processes in complex porous media. Realistic pore-spaces may be characterized by singular regions, i.e., isolated subnetworks that do not connect inlet and outlet, resulting from unconnected porosity or multiphase configurations. The robust identification of these features is critical for the characterization of network topology and the solution of the set of linear equations of flow and transport. We propose a robust method based on singular value decomposition to solve for network flow and locate isolated subnetworks simultaneously. The method's performance is demonstrated for networks of different complexity.  
\end{abstract}


\begin{highlights}
\item Singular value decomposition (SVD) is an efficient and robust method to solve for single-phase flow in heterogeneous networks with isolated clusters
\item SVD can be used to locate singular points and clusters
\item Singular value decomposition is robust, and allows for determining the percolation threshold in heterogeneous networks.
\end{highlights}

\end{frontmatter}
\section{Introduction} \label{sec:PNM}
Conceptualization of the media as a pore network (rather than a bundle of capillaries) goes back to the work of \citet{Fatt1956} and \citet{broadbent1957percolation}, further advanced by the seminal works of \citet{wilkinson1983invasion, wilkinson1986percolation, lenormand1983mechanisms}. In its simplest form, a porous medium is conceptualized as a lattice of conductors, disregarding the pore shape. More complex (and realistic) pore network models relate the bond geometry to their conductance \citep{Sahimi2011} and map out the pore structure obtained from pore space partitioning \citep{Blunt2017}. 

Depending on the 
complexity of the pore space, the resulting pore network may be characterized by subnetworks or clusters that are hydraulically isolated, that is, clusters that are not connecting the inlet and outlet of the pore network \citep{Raoof2010}. Isolated sites cause numerical problems for the solution of flow in networks because they lead to singular system matrices. 
For example, the dilution of a pore network \citep{hunt2017}
by randomly assigning zero conductances to connections between sites may lead to isolated subnetworks. This is a general issue for flow problems in media characterized by isolated clusters.
Such media can result from the dilution of regular networks \citep{hunt2017}, from networks patterns generated by invasion percolation, or from pore networks derived directly from image analysis of porous and fractured media characterized by unconnected porosity. The singularity issue can be addressed by preprocessing of the network image and removing isolated clusters before the flow solution. Algorithms for the cluster identification include the methods of \citet{sheppard1999invasion} and \citet{Raoof2010} and \citet{guo2021}.      

In this note, we propose a robust algorithm based on singular value decomposition (SVD) of the system matrix in order to both solve the linear flow problem and identify isolated clusters in the network.  
SVD decomposes a matrix into the product of a diagonal and two orthonormal matrices \citep{golub2013matrix}. It is used, for example, to determine the pseudoinverse of a singular matrix and calculate low-rank approximations of a matrix. SVD finds broad applications in data-driven science and engineering \citep{Brunton_Kutz_2019, bisgard2020analysis}. The idea here is to use SVD to identify nodes belonging to isolated clusters by solving for single-phase flow in the PNM using the pseudoinverse of the Laplacian matrix. The internal nodes of zero pressure belong to isolated clusters.  The proposed method is general and applies to the identification of isolated clusters in networks. 

In this technical note, we first evaluate the accuracy of the SVD pseudoinverse approximation, then demonstrate its robustness, and finally present a running-time evaluation example.

\section{Methods\label{sec:methods}}
%
\subsection{Graph Laplacian}
The graph Laplacian is a representation of the graph underlying the PNM \cite{bapat2010graphs}. The Laplacian is weighted by the conductances of its bonds of the graph. 
The conductance or weight matrix $\mathbf G$ contains the conductances between $G_{ij}$ between sites $i$ and $j$. If all non-zero conductances are unity, $\mathbf G$ is equal to the adjacency matrix of the graph. The weighted degree matrix is a diagonal matrix that contains the sum of the conductances of the bonds adjacent to a node, that is, $D_{ij} = \delta_{ij} \sum_{[ki]} G_{ik}$, where $\sum_{[ki]}$ denotes summation over all sites $k$ that are connected to site $i$. With these definitions, the weighted graph Laplacian is given by $\mathbf L' = \mathbf D - \mathbf G$. The continuity equation for single-phase flow through the PNM can then be written as 
\begin{align}
\label{eq:matrix_Ohm}
\mathbf L' \cdot \mathbf P = \mathbf 0,
\end{align}
where $\mathbf P$ is the vector of pressures at all sites of
the network. Following Poiseuille law \citep{Poiseuille1839}, the flow rate between sites $i$ and $j$ is 
\begin{align}
    Q_{ij} = G_{ij} (P_i - P_j),  
\end{align}
which is a statement of momentum conservation. Notice that the pressure at the unconnected site does not affect the flow field as $G_{ij}=0$ on all bonds in the unconnected cluster periphery. Also note, that $Q_{ij} = - Q_{ji}$.  
Continuity, or mass conservation, is expressed by
\begin{align}
\label{eq:mass}
    \sum_{[ji]} Q_{ij} = 0.
\end{align}
The latter is equivalent to Eq.~\eqref{eq:matrix_Ohm}. The flow rates at sites are defined by
\begin{align}
Q_{i}= \frac{1}{2}\sum_{[ji]} |Q_{ij}|.
\end{align}
From the mass conservation statement~\eqref{eq:mass}, it follows that the graph Laplacian has the eigenvector $(1,\dots,1)$ with the eigenvalue zero, which is the trivial solution of equal pressure at all sites. Thus, the determinant of the graph Laplacian is zero, i.e., $\mathbf L'$ is not invertible. The multiplicity of the zero eigenvalue is equal to the number of isolated clusters in the graph, or equivalently, the rank of the graph Laplacian is equal to the number of sites minus the number of connected components \citep{bapat2010graphs}. Note that the sites belonging to an isolated cluster do not contribute separately to this count, that is, an isolated cluster, even though is may contain multiple sites, counts as one.  

We seek a solution to Eq.~\eqref{eq:matrix_Ohm} for prescribed pressure at the inlet and outlet nodes. To this end, we define the modified graph Laplacian $\mathbf L$ such that $L_{ij} = \delta_{ij}$ for all sites $i,j$ that are in the inlet or outlet boundaries and $L_{ij} = L'_{ij}$ for all other sites. With this definition, the boundary value problem can be written as 
\begin{equation} \label{eq:matrix_Ohm2}
{\mathbf L} \cdot \mathbf P = \mathbf b,
\end{equation}
where the vector $\mathbf b$ on the right side of Eq.~\eqref{eq:matrix_Ohm2} contains the boundary conditions with $b_{i} = p_0$ for the sites $j$ at the inlet boundary and $b_i = 0$ else. For networks that consist of a single connected cluster, the modified Laplacian $\mathbf L$ is an invertible (non-singular) matrix. The trivial zero eigenvalue proper to the graph Laplacian is eliminated by specifying non-trivial boundary conditions. Only in the presence of isolated clusters has the modified graph Laplacian $\mathbf L$ a zero eigenvalue and thus is not invertible. This can be interpreted by the example of a single isolated internal site of the network. In this case, a row in the modified Laplacian is zero, and thus, its determinant is zero as well. 
In order to solve the flow equation~\eqref{eq:matrix_Ohm2}, and to determine the sites belonging to isolated regions, we propose to use the singular value decomposition (SVD) of the modified graph Laplacian. In the following, we provide a brief summary of SVD.      
\subsection{Singular Value Decomposition} \label{sec:SVD}
The singular value decomposition of the matrix $\mathbf L$ is given by the product of two orthonormal matrices $\mathbf U$ and $\mathbf V$ and a diagonal matrix $\boldsymbol \Sigma$ with positive real entries $\sigma_i$ as \citep{golub2013matrix}
\begin{equation} \label{eq:SVD}
\mathbf L=\mathbf U \boldsymbol \Sigma \mathbf V^T,
\end{equation}
The entries $\sigma_i$ are called the singular values of $\mathbf L$. The number of non-zero $\sigma_i$ is defined as the rank of the matrix $\mathbf L$. The columns of $\mathbf V$ (termed the right singular vectors of $\mathbf L$), and the columns of $\mathbf U$ (the left singular vectors), always form an orthogonal set with no assumptions on $\mathbf L$. If $\mathbf L$ is an invertible matrix, its inverse is given by
\begin{equation} 
\label{eq:inverse}
\mathbf L^{-1}= \mathbf V \boldsymbol \Sigma ^{-1} \mathbf U^T.
\end{equation} 

If $\mathbf L$ is a singular matrix (i.e., at least one $\sigma_i=0$), its pseudo-inverse $\mathbf L^+$ is determined as follows \citep{golub2013matrix}. First, the pseudo-inverse $\boldsymbol \Sigma^+$ of the diagonal matrix $\boldsymbol \Sigma$ is determined by inverting every non-zero element on the diagonal and leaving the zero elements unchanged. In the numerical calculations, elements are inverted if they are larger than a given threshold and set equal to zero if they are smaller. The pseudo-inverse $\mathbf L^+$ is then given by 
\begin{align}
\label{eq:SVD_inverse}
    \mathbf L^+ = \mathbf V \mathbf \Sigma ^{+} \mathbf U^T
\end{align}
Once the pseudo-inverse is obtained, the pressure distribution is determined by a matrix multiplication as
\begin{equation} \label{eq:pressure}
\mathbf P = \mathbf L^+ \mathbf b. 
\end{equation} 
The sites $i$ belonging to isolated clusters have zero pressure $P_i = 0$ because they are not connected to inlet and outlet. This approach does not require matrix preconditioning and can be used for singular and invertible matrices. At the same time, it is robust and accurate.

\subsection{Pore network from a partially saturated medium} \label{sec:meth_image}
\begin{figure}
 \includegraphics[width=\textwidth]{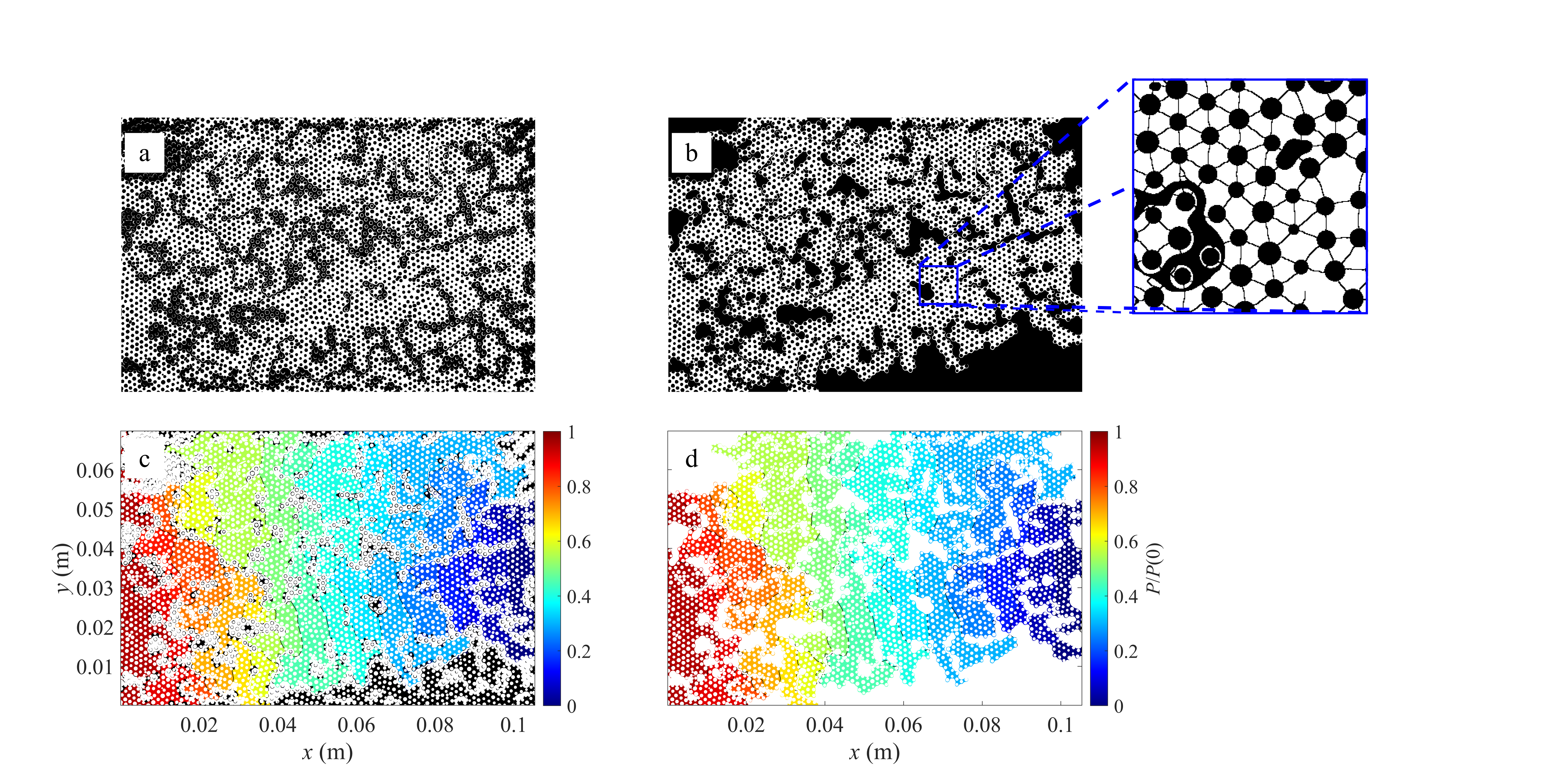}
 \caption{a) Binary image of the liquid phase (white) configuration during multi-phase flow through a milifluidic device (image provided by Yves Méheust). b) Binary image of the processed image after removing the disconnected and non-percolated liquid phase. In the small plot in b), the partitioned liquid-filled pore space in a zoom-in section (blue square), pore throats, and stagnant matrix are in black. c) Pressure distribution in the pores using SVD to solve the non-processed image (a), singular regions are assigned with a value of 0, P values smaller than 0.1\%P(0), corresponding to the singular sites, are colored in black.
 d) Pressure distribution (not using SVD) in the pores of the processed image (b).} \label{fig:media} \end{figure}

We compare the proposed SVD-based method to a method that uses first a pre-processing step in order to identify and remove isolated clusters, and then solve the flow problem by standard inversion of the system matrix. To this end, we consider a pore network characterized by isolated regions that is obtained by pore space partitioning of an image of the fluid phase of a two-dimensional millifluidic device under partially saturated single-phase flow conditions.
Figure~\ref{fig:media} shows the fluid phase (in white) distribution in a milifluidic device before (Fig.~\ref{fig:media}a) and after (Fig.~\ref{fig:media}b) processing the image to exclude the non-percolated disconnected fluid clusters. 
The experimental setup and device specification are described in \citet{JM2017}. 
In this work, we only use the image to construct a relevant pore network and do not address the physics arising from the experiment.
The original image (Fig.~\ref{fig:media}a) contains many disconnected regions of different sizes. 
The small regions are formed by wetting liquid phase films surrounding the grains, while the larger clusters are formed by the entrapment of the liquid phase by the non-wetting gas phase. 
The processed image (Fig.~\ref{fig:media}b) does not include isolated fluid clusters. 
The pre-processing of the pixelized image was conducted using the MATLAB image analysis toolbox \citep{MATLAB} using the 8-connectivity criterion to identify isolated clusters.

The pore-scale partitioning, both in the original and pre-processed images, is performed following the methodology presented in \citep{ben2024Image}. 
This method uses information about the curvatures of the distance map (obtained from the binary image) to separate distinct pores (small plot in Fig.~\ref{fig:media}b) and to evaluate the pore body and throat sizes. 
The construction of the detailed network follows the methodology presented in \citep{ben2024network}. 
A parabolic parameterization of the pore geometry is assumed to evaluate the bond conductivities in the resulting pore network models. 
For laminar flow, the bond conductance is evaluated by the harmonic sum of the local conductivities along the pore channel with variable width by using \citet{Boussinesq1868} equation for rectangular apertures. 
The pore sizes and conductance distributions and additional information about the milifluidic device and pore partitioning are provided in Figure SF1 in the supplementary material.
The results of this analysis are presented in the subsection \ref{sec:image} below.

\subsection{Diluted pore networks} \label{sec:meth_network}
We use the SVD method to solve the steady flow equation on two types of diluted networks based on two-dimensional hexagonal and square lattices at different degrees of dilution. In short, networks are diluted by randomly assigning zero conductances to connections between sites. The network properties, such as the the probability of a cluster connecting inlet and outlet, can be determined by percolation theory based on the occupation probability ($p$) of either lattice bonds or lattice sites and their conductances \citep{berkowitz1998percolation, broadbent1957percolation, hunt2017, stauffer2018introduction}. We compare the percolation properties determined from the SVD-based flow solution to theoretical values from percolation theory. 

We consider two types of diluted lattices based on two-dimensional hexagonal 2D lattices with coordination number $Z_s=3$ and a two-dimensional square 2D lattice with coordination number $Z_s=4$. 
For simplicity, we use a uniform conductance field ($G_{ij}=1$ for all $i,j$), unit bond length of $\ell = 1$, $N = 100$ sites along each face of the lattices, and a unit pressure gradient $\Delta P/L$. 
The domain length is $L = N \ell$ for the square lattice, and $L ={3 N \ell}/{4}$ for the hexagonal lattice, which is oriented with the long diagonal in the main flow direction. 
The total number of sites in the regular network is denoted by $N_s$.

Starting from a regular lattice with coordination number $Z_s$, the network is diluted to obtain a target occupation probability 
\begin{align}
p_b = \frac{\langle Z \rangle}{Z_{s}},
\end{align}
where $\langle Z \rangle$ is the mean coordination number of the diluted network. 
The dilution method applied here is a simplification of the protocol described in \citep{Raoof2010}. 
Each bond is randomly assigned a dilution or elimination number between zero and one. 
Then, all bonds with an elimination number larger than $p_b$ are assigned zero conductivity. 
In this way, $N_R = 50$ network realizations are generated for each $p_b$ value. 
Then, for each realization, the flow equation \eqref{eq:matrix_Ohm2} is solved using the pseudoinverse given by Eq.~\ref{eq:SVD_inverse}. 

We determine for each target $p_b$ value the number of realizations for which a percolating cluster prevails. 
A percolating cluster exists if the permeability defined by Eq.~\eqref{eq:perm} is larger than 0. 
The percolation frequency $p_p$ for a given $p_b$ is defined as the ratio of the number $N_p$ of realizations with percolating clusters and the total number $N_R$ of realizations,
\begin{align}
    p_p = \frac{N_p}{N_R}. 
\end{align}
For $p_b$ smaller than a critical value $p_c$, the bond percolation threshold, there is no percolating cluster, that is, $p_p = 0$. 
The theoretical random bond percolation threshold for an infinite hexagonal lattice is $p_c = 1-2 \sin({\pi}/{18})\approx0.6527$, and for an infinite square lattice it is $p_c = \frac{1}{2}$ \citep{hunt2017}. 

These synthetic media are used to evaluate the robustness, that is, the ability of the SVD method to solve the flow equation of a media with many singular points and isolated clusters. The results of this analysis are presented in the subsection \ref{sec:network} below.

\section{Demonstration and Validation} \label{sec:validation}
In this section, we demonstrate and validate the proposed SVD approach.
First, in subsection \ref{sec:image}, we evaluate the accuracy of the method.
Then, in section \ref{sec:network}, we demonstrate the robustness of the method.

\subsection{Pore network from a partially saturated medium} \label{sec:image}
In this subsection, we consider a pore network obtained by pore space partitioning of an image of the fluid phase of a two-dimensional millifluidic device under partially saturated single-phase flow conditions (see subsection \ref{sec:meth_image}).
The results are compared in terms of flow rate and pressure statistics to assess the accuracy of the proposed method. 
The total flow rate ($Q_m$) is defined by the sum of the flow rates at the network's inlet sites. 
\begin{align}
\label{eq:Qm}
    Q_m = {\sum_i Q_i},
\end{align}

We determine the probability density function (PDF) of the flow rate at the sites through areal sampling from the network flow simulations and image pore space partitioning 
\begin{align}
\label{eq:fQ}
f(Q_k) = \frac{\sum_i \mathbb I(Q_i,Q_k) A_i}{\Delta Q_k\sum_i A_i},
\end{align}
where $A_i$ [L$^2$] is the area of pore $i$, and $Q_i$ is the flow rate at the $i$-th site. The indicator function $\mathbb I$ is one if the flow rate $Q_i$ is within the interval $(Q_k, Q_k + \Delta Q_k]$ and zero otherwise. The pore area ($A_i$) is evaluated by summing the number of pixels in the image belonging to each pore and multiplying it by a single pixel area ($10^{-3}$ mm$^2$). 

The distribution of the pressure at the sites is determined in analogy to the distribution of flow rates
\begin{align}
\label{eq:fP}
f(P_k) = \frac{\sum_i \mathbb I(P_i,P_k) A_i}{\Delta P_k\sum_i A_i},
\end{align}

For the network constructed from the original image, the flow equation \eqref{eq:matrix_Ohm2} is solved using the pseudoinverse \eqref{eq:SVD_inverse} obtained from SVD, which gives the pressures at all sites through Eq.\eqref{eq:pressure}. 
For the network obtained from the pre-processed image, the flow equation is solved by direct inversion of Eq.~\ref{eq:matrix_Ohm2} and by using the pseudoinverse from SVD.   

\begin{figure} 
 \includegraphics[width=\textwidth]{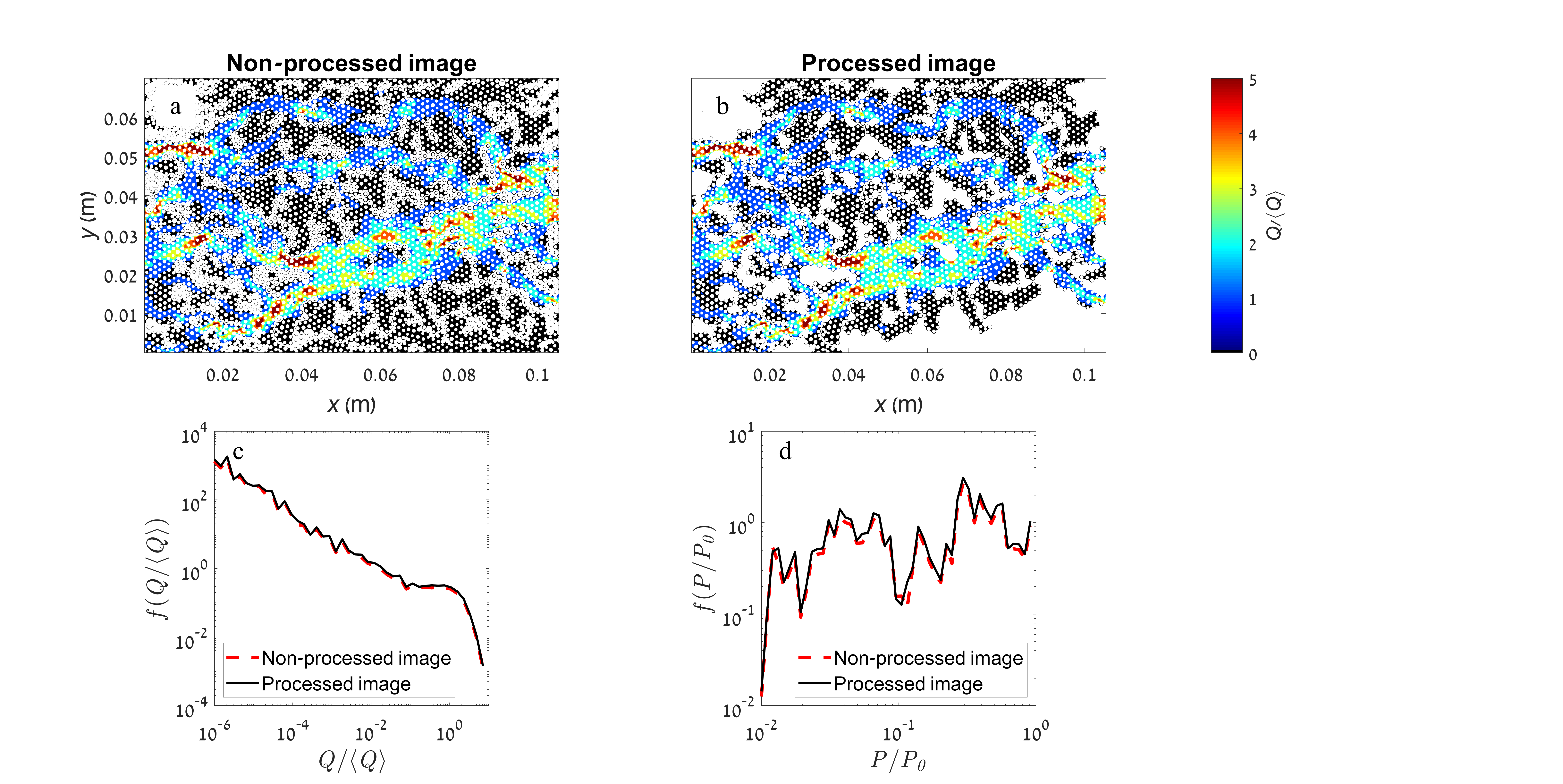}
 \caption{Top panel: Flow rates at the network sites, for 
 a) the network obtained from the original image, evaluated using SVD, 
 and b) pre-processed image, using direct matrix inverse. Flow rates lower than 1\% of the mean flow rate are colored in black.
 Bottom panel: Area-weighted probability density functions of 
 c) the flow rate at the sites 
 and d) the pressure at the sites for the networks obtained from the original and pre-processed images. 
 The flow rates are normalized by the pre-processed image's mean flow rate ($\langle Q \rangle$).}     
 \label{Qs_pdf_SVD} 
 \end{figure}

Figure~\ref{Qs_pdf_SVD}a and b show the spatial distribution of flow rates in the pore networks obtained from the original and pre-processed images. 
The singular disconnected regions attain a zero pressure value. 
In the network resulting from the original image, the solution using SVD assigns zero pressure to the isolated clusters. 
The clusters obtained from the SVD analysis on the original image coincide with those identified from the pre-processing step of the binary image. 
The spatial flow rate distribution in the two networks is essentially identical. 
This is confirmed by the probability density function of bond flow rates and pressures for both networks, shown in the bottom panel of Figure~\ref{Qs_pdf_SVD}. 
Moreover, the device flow rate $Q_m$ defined in Eq.~\eqref{eq:Qm} for the networks from the original and pre-processed images differ by about $0.2\%$ further confirming the performance and accuracy of the proposed SVD-based method. 

In order to assess whether this difference is a result of the solution of the flow equation by different methods (SVD versus direct inversion) or due to the pre-processing, we use the SVD method to solve for the flow on the pre-processed image, despite the pre-processed image does not have any singularities. The difference in the total flow rates obtained by the two methods is smaller than $10^{-11}$, that is, they are essentially identical. This implies that there is a slight difference in the identification of isolated clusters by the SVD method on the pore network extracted from the direct image and by the connectivity-based method on the pixel image. In other words, the connected percolating clusters in the pore networks extracted from the direct and pre-processed images are not exactly the same. Thus, the SVD-based method provides an accurate method for the identification of isolated clusters and solution of single phase flow in the network obtained from the direct image without the need for pre-processing, which may introduce some inaccuracies.

\subsection{Diluted pore networks} \label{sec:network}

In this section, we evaluate the ability of the SVD-based method to solve the flow equation and identify isolated clusters at bond occupation probabilities close to the percolation threshold (see subsection \ref{sec:meth_network}). 

In addition to the percolation thresholds, the results for this media, are compared in terms of relative conductivity, which is defined in terms of the total flow rate relative to the total flow rate in a corresponding regular network as
\begin{align}
\label{eq:perm}
    k_r = \frac{Q_m}{Q_m^{(0)}},
\end{align}
where the superscript $0$ denotes to the flow rate for the corresponding regular network.

\begin{figure}
 \includegraphics[width=\textwidth]{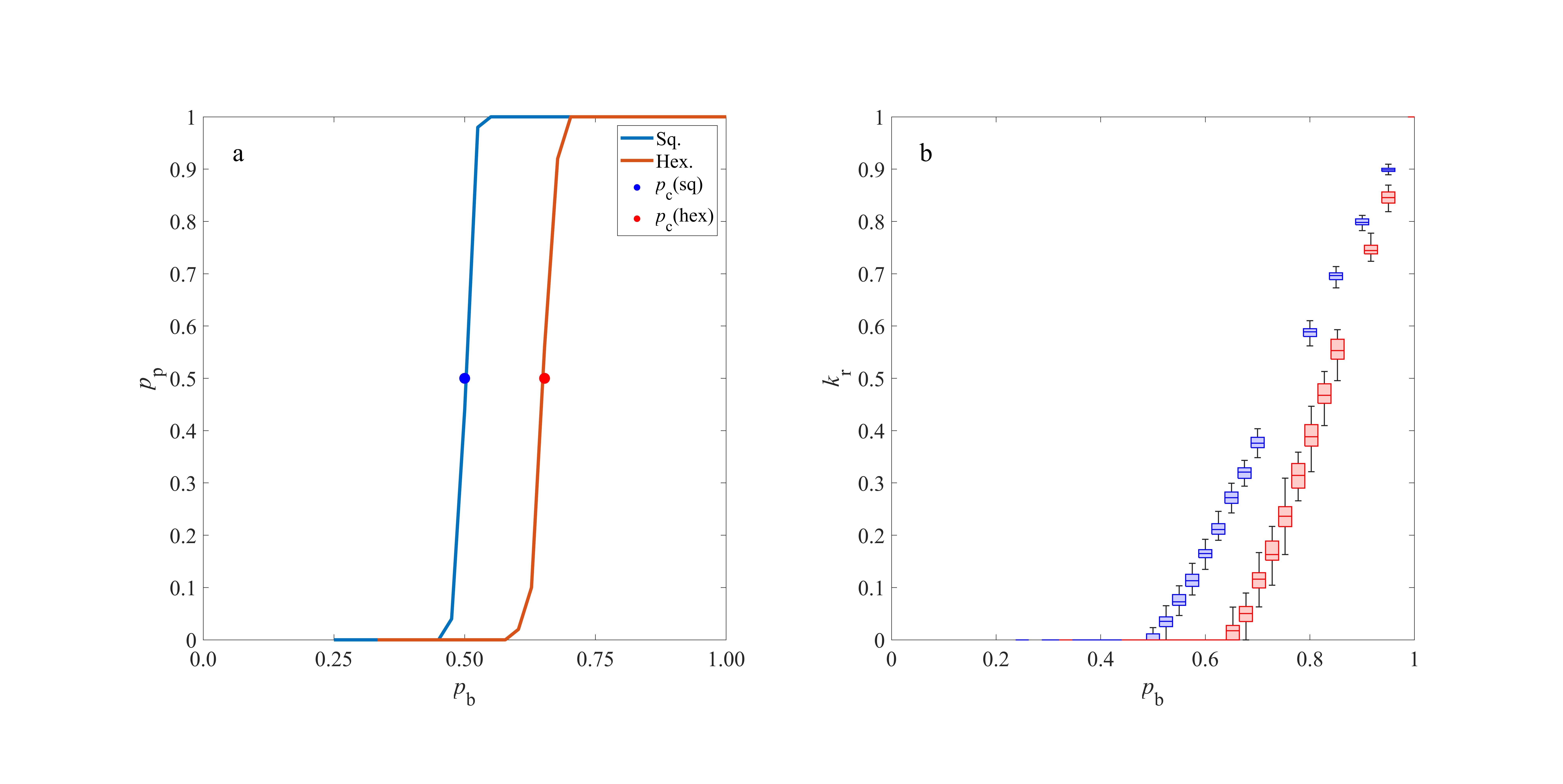}
 \caption{ The effect of the occupation probability ($p_b$) on: a) the percolation probability ($p_p$)
 and b) the relative lattice permeability ($k_r$) for square (blue) and hexagonal (red) lattices. 
 Lattices with $100$ sites in each main direction and $50$ realizations for each $p_b$ value are used for this analysis. 
 The symbols in a) denote the theoretical values of $p_c$ for the square and hexagonal lattices.}\label{fig:p_p}  
 \end{figure}

From the SVD-based numerical flow simulations in the diluted networks, the percolation threshold $p_c$ is estimated as the bond occupation probability $p_b$ for which $50\%$ of the lattice realizations percolate \citep{Friedman1998}.
As shown in Figure \ref{fig:p_p}a, the numerical values obtained for $p_c$ are in very good agreement with the theoretical values both for the hexagonal ($p_c \approx0.6527$) and square ($p_c = 0.5$) lattices. 
A slight deviation of the numerical from the theoretical values can be attributed to the finite size of the lattices with $N = 100$, and the finite number of realizations with $N_R=50$. 
Figure \ref{fig:p_p}b shows the dependence of permeability on $p_b$. 
For $p_b < p_c$, permeability tends to zero, then for $p_b > p_c$ it increases monotonically toward $k_r = 1$, which is the value for the non-diluted lattices. 
For $p_b$ far from $p_c$, the evolution is approximately linear. These behaviors are in accordance with the literature \citep{Kirkpatrick1973, Friedman1995}. 
The variability of the $k_r$ values between realizations, illustrated by the width of the error bars, is not linear, with zero variance for $p_b = 1$ and $p_b \ll p_c$ and a large variance for intermediate values. These results demonstrate the robustness of the proposed SVD method.

\section{Conclusion} \label{sec:conc}
We have shown that SVD is a robust, easy-to-implement, and accurate method to identify sites belonging to isolated clusters in heterogeneous networks. 
Isolated clusters render the modified graph Laplacian non-invertible. 
Thus, we use the pseudoinverse determined from SVD to solve for single-phase flow in the heterogeneous networks, which assigns zero pressure to sites belonging to isolated clusters. 
The SVD method is demonstrated and validated for pore network models, which are characterized by singular points and isolated clusters. The time complexity for algorithms to identify isolated clusters such as the one proposed in \cite{sheppard1999invasion} is of order $N \ln N$, where $N$ is the number of sites, while the time complexity for matrix inversion is of the order of $N^3$. The proposed SVD method, which provides at the same time the flow solution and delineates the isolated clusters, is of order $N^3$. This shows that SVD provides a robust alternative to combined algorithms that identify isolated clusters and solve for single phase flow sequentially. 

 \section*{Acknowledgments}
  The authors thank Yves M\'eheust for the image of the milifluidic device. I.B.N, M.D. acknowledge funding from the European Union’s Horizon 2020 research and innovation programme under the Marie Sklodowska-Curie grant agreement No. HORIZON-MSCA-2021-PF-01 (USFT). I.B.N. , J.J.H. and M.D. acknowledge the support of the Spanish Research Agency (10.13039/501100011033), Spanish Ministry of Science, Innovation and Universities through grant HydroPore II PID2022-137652NB-C41. J.J.H. acknowledges the support of the Spanish Research Agency (10.13039/501100011033), the Spanish Ministry of Science and Innovation and the European Social Fund ``Investing in your future'' through the ``Ram\'on y Cajal'' fellowship (RYC-2017-22300). 

 \section{Open Research}
No data was used in this research.

\bibliographystyle{elsarticle-harv} 
\biboptions{authoryear}
\bibliography{sample}

\begin{thebibliography}{26}
\expandafter\ifx\csname natexlab\endcsname\relax\def\natexlab#1{#1}\fi
\providecommand{\url}[1]{\texttt{#1}}
\providecommand{\href}[2]{#2}
\providecommand{\path}[1]{#1}
\providecommand{\DOIprefix}{doi:}
\providecommand{\ArXivprefix}{arXiv:}
\providecommand{\URLprefix}{URL: }
\providecommand{\Pubmedprefix}{pmid:}
\providecommand{\doi}[1]{\href{http://dx.doi.org/#1}{\path{#1}}}
\providecommand{\Pubmed}[1]{\href{pmid:#1}{\path{#1}}}
\providecommand{\bibinfo}[2]{#2}
\ifx\xfnm\relax \def\xfnm[#1]{\unskip,\space#1}\fi
\bibitem[{Bapat(2010)}]{bapat2010graphs}
\bibinfo{author}{Bapat, R.B.}, \bibinfo{year}{2010}.
\newblock \bibinfo{title}{Graphs and matrices}. volume~\bibinfo{volume}{27}.
\newblock \bibinfo{publisher}{Springer}.
\bibitem[{Ben-Noah et~al.(2024a)Ben-Noah, Hidalgo and Dentz}]{ben2024Image}
\bibinfo{author}{Ben-Noah, I.}, \bibinfo{author}{Hidalgo, J.J.}, \bibinfo{author}{Dentz, M.}, \bibinfo{year}{2024}a.
\newblock \bibinfo{title}{Efficient pore space characterization based on the curvature of the distance map}.
\newblock \href{http://arxiv.org/abs/2403.12591}{{\tt arXiv:2403.12591}}.
\bibitem[{Ben-Noah et~al.(2024b)Ben-Noah, Hidalgo and Dentz}]{ben2024network}
\bibinfo{author}{Ben-Noah, I.}, \bibinfo{author}{Hidalgo, J.J.}, \bibinfo{author}{Dentz, M.}, \bibinfo{year}{2024}b.
\newblock \bibinfo{title}{Evaluation of different network models in variably saturated media}.
\newblock \href{http://arxiv.org/abs/2403.13519}{{\tt arXiv:2403.13519}}.
\bibitem[{Berkowitz and Ewing(1998)}]{berkowitz1998percolation}
\bibinfo{author}{Berkowitz, B.}, \bibinfo{author}{Ewing, R.P.}, \bibinfo{year}{1998}.
\newblock \bibinfo{title}{Percolation theory and network modeling applications in soil physics}.
\newblock \bibinfo{journal}{Surveys in Geophysics} \bibinfo{volume}{19}, \bibinfo{pages}{23--72}.
\bibitem[{Bisgard(2020)}]{bisgard2020analysis}
\bibinfo{author}{Bisgard, J.}, \bibinfo{year}{2020}.
\newblock \bibinfo{title}{Analysis and linear algebra: the singular value decomposition and applications}. volume~\bibinfo{volume}{94}.
\newblock \bibinfo{publisher}{American Mathematical Soc.}
\bibitem[{Blunt(2017)}]{Blunt2017}
\bibinfo{author}{Blunt, M.J.}, \bibinfo{year}{2017}.
\newblock \bibinfo{title}{Multiphase flow in permeable media: A pore-scale perspective}.
\newblock \bibinfo{publisher}{Cambridge university press}.
\bibitem[{Boussinesq(1868)}]{Boussinesq1868}
\bibinfo{author}{Boussinesq, J.}, \bibinfo{year}{1868}.
\newblock \bibinfo{title}{M{\'e}moire sur l’influence des frottements dans les mouvements r{\'e}guliers des fluids}.
\newblock \bibinfo{journal}{Journal de math{\'e}matiques pures et appliqu{\'e}es} \bibinfo{volume}{13}, \bibinfo{pages}{377--424}.
\bibitem[{Broadbent and Hammersley(1957)}]{broadbent1957percolation}
\bibinfo{author}{Broadbent, S.R.}, \bibinfo{author}{Hammersley, J.M.}, \bibinfo{year}{1957}.
\newblock \bibinfo{title}{Percolation processes: I. crystals and mazes}, in: \bibinfo{booktitle}{Mathematical proceedings of the Cambridge philosophical society}, \bibinfo{organization}{Cambridge University Press}. pp. \bibinfo{pages}{629--641}.
\bibitem[{Brunton and Kutz(2019)}]{Brunton_Kutz_2019}
\bibinfo{author}{Brunton, S.L.}, \bibinfo{author}{Kutz, J.N.}, \bibinfo{year}{2019}.
\newblock \bibinfo{title}{Data-Driven Science and Engineering: Machine Learning, Dynamical Systems, and Control}.
\newblock \bibinfo{publisher}{Cambridge University Press}.
\bibitem[{Fatt(1956)}]{Fatt1956}
\bibinfo{author}{Fatt, I.}, \bibinfo{year}{1956}.
\newblock \bibinfo{title}{The network model of porous media}.
\newblock \bibinfo{journal}{Transactions of the AIME} \bibinfo{volume}{207}, \bibinfo{pages}{144--177}.
\bibitem[{Friedman and Seaton(1998)}]{Friedman1998}
\bibinfo{author}{Friedman, S.P.}, \bibinfo{author}{Seaton, N.}, \bibinfo{year}{1998}.
\newblock \bibinfo{title}{Percolation thresholds and conductivities of a uniaxial anisotropic simple-cubic lattice}.
\newblock \bibinfo{journal}{Transport in porous media} \bibinfo{volume}{30}, \bibinfo{pages}{241--250}.
\bibitem[{Friedman et~al.(1995)Friedman, Zhang and Seaton}]{Friedman1995}
\bibinfo{author}{Friedman, S.P.}, \bibinfo{author}{Zhang, L.}, \bibinfo{author}{Seaton, N.A.}, \bibinfo{year}{1995}.
\newblock \bibinfo{title}{Gas and solute diffusion coefficients in pore networks and its description by a simple capillary model}.
\newblock \bibinfo{journal}{Transport in porous media} \bibinfo{volume}{19}, \bibinfo{pages}{281--301}.
\bibitem[{Golub and Van~Loan(2013)}]{golub2013matrix}
\bibinfo{author}{Golub, G.H.}, \bibinfo{author}{Van~Loan, C.F.}, \bibinfo{year}{2013}.
\newblock \bibinfo{title}{Matrix computations}.
\newblock \bibinfo{publisher}{JHU press}.
\bibitem[{Guo et~al.(2021)Guo, Yang, Jia, Tao, Xu, Dong and Liu}]{guo2021}
\bibinfo{author}{Guo, X.}, \bibinfo{author}{Yang, K.}, \bibinfo{author}{Jia, H.}, \bibinfo{author}{Tao, Z.}, \bibinfo{author}{Xu, M.}, \bibinfo{author}{Dong, B.}, \bibinfo{author}{Liu, L.}, \bibinfo{year}{2021}.
\newblock \bibinfo{title}{A new method of central axis extracting for pore network modeling in rock engineering}.
\newblock \bibinfo{journal}{Geofluids} \bibinfo{volume}{2021}, \bibinfo{pages}{1--20}.
\bibitem[{Hunt and Sahimi(2017)}]{hunt2017}
\bibinfo{author}{Hunt, A.G.}, \bibinfo{author}{Sahimi, M.}, \bibinfo{year}{2017}.
\newblock \bibinfo{title}{Flow, transport, and reaction in porous media: Percolation scaling, critical-path analysis, and effective medium approximation}.
\newblock \bibinfo{journal}{Reviews of Geophysics} \bibinfo{volume}{55}, \bibinfo{pages}{993--1078}.
\bibitem[{Jim{\'e}nez-Mart{\'\i}nez et~al.(2017)Jim{\'e}nez-Mart{\'\i}nez, Le~Borgne, Tabuteau and M{\'e}heust}]{JM2017}
\bibinfo{author}{Jim{\'e}nez-Mart{\'\i}nez, J.}, \bibinfo{author}{Le~Borgne, T.}, \bibinfo{author}{Tabuteau, H.}, \bibinfo{author}{M{\'e}heust, Y.}, \bibinfo{year}{2017}.
\newblock \bibinfo{title}{Impact of saturation on dispersion and mixing in porous media: Photobleaching pulse injection experiments and shear-enhanced mixing model}.
\newblock \bibinfo{journal}{Water Resources Research} \bibinfo{volume}{53}, \bibinfo{pages}{1457--1472}.
\bibitem[{Kirkpatrick(1973)}]{Kirkpatrick1973}
\bibinfo{author}{Kirkpatrick, S.}, \bibinfo{year}{1973}.
\newblock \bibinfo{title}{Percolation and conduction}.
\newblock \bibinfo{journal}{Reviews of modern physics} \bibinfo{volume}{45}, \bibinfo{pages}{574}.
\bibitem[{Lenormand et~al.(1983)Lenormand, Zarcone and Sarr}]{lenormand1983mechanisms}
\bibinfo{author}{Lenormand, R.}, \bibinfo{author}{Zarcone, C.}, \bibinfo{author}{Sarr, A.}, \bibinfo{year}{1983}.
\newblock \bibinfo{title}{Mechanisms of the displacement of one fluid by another in a network of capillary ducts}.
\newblock \bibinfo{journal}{Journal of Fluid Mechanics} \bibinfo{volume}{135}, \bibinfo{pages}{337--353}.
\bibitem[{Poiseuille(1839)}]{Poiseuille1839}
\bibinfo{author}{Poiseuille, J.L.M.}, \bibinfo{year}{1839}.
\newblock \bibinfo{title}{Recherches sur les causes du mouvement du sang dans les vaisseaux capillaires}. volume~\bibinfo{volume}{7}.
\newblock \bibinfo{publisher}{Impr. royale}.
\bibitem[{Raoof and Hassanizadeh(2010)}]{Raoof2010}
\bibinfo{author}{Raoof, A.}, \bibinfo{author}{Hassanizadeh, S.M.}, \bibinfo{year}{2010}.
\newblock \bibinfo{title}{A new method for generating pore-network models of porous media}.
\newblock \bibinfo{journal}{Transport in porous media} \bibinfo{volume}{81}, \bibinfo{pages}{391--407}.
\bibitem[{Sahimi(2011)}]{Sahimi2011}
\bibinfo{author}{Sahimi, M.}, \bibinfo{year}{2011}.
\newblock \bibinfo{title}{Flow and transport in porous media and fractured rock: from classical methods to modern approaches}.
\newblock \bibinfo{publisher}{John Wiley \& Sons}.
\bibitem[{Sheppard et~al.(1999)Sheppard, Knackstedt, Pinczewski and Sahimi}]{sheppard1999invasion}
\bibinfo{author}{Sheppard, A.P.}, \bibinfo{author}{Knackstedt, M.A.}, \bibinfo{author}{Pinczewski, W.V.}, \bibinfo{author}{Sahimi, M.}, \bibinfo{year}{1999}.
\newblock \bibinfo{title}{Invasion percolation: new algorithms and universality classes}.
\newblock \bibinfo{journal}{Journal of Physics A: Mathematical and General} \bibinfo{volume}{32}, \bibinfo{pages}{L521}.
\bibitem[{Stauffer and Aharony(2018)}]{stauffer2018introduction}
\bibinfo{author}{Stauffer, D.}, \bibinfo{author}{Aharony, A.}, \bibinfo{year}{2018}.
\newblock \bibinfo{title}{Introduction to percolation theory}.
\newblock \bibinfo{publisher}{CRC press}.
\bibitem[{{The~MathWorks~Inc.}(2022)}]{MATLAB}
\bibinfo{author}{{The~MathWorks~Inc.}}, \bibinfo{year}{2022}.
\newblock \bibinfo{title}{Matlab version: 9.13.0 (r2022b)}.
\newblock \URLprefix \url{https://www.mathworks.com}.
\bibitem[{Wilkinson(1986)}]{wilkinson1986percolation}
\bibinfo{author}{Wilkinson, D.}, \bibinfo{year}{1986}.
\newblock \bibinfo{title}{Percolation effects in immiscible displacement}.
\newblock \bibinfo{journal}{Physical Review A} \bibinfo{volume}{34}, \bibinfo{pages}{1380}.
\bibitem[{Wilkinson and Willemsen(1983)}]{wilkinson1983invasion}
\bibinfo{author}{Wilkinson, D.}, \bibinfo{author}{Willemsen, J.F.}, \bibinfo{year}{1983}.
\newblock \bibinfo{title}{Invasion percolation: a new form of percolation theory}.
\newblock \bibinfo{journal}{Journal of physics A: Mathematical and general} \bibinfo{volume}{16}, \bibinfo{pages}{3365}.

\end{thebibliography}

\end{document}


\begin{frontmatter}

\title{Singular Value Decomposition for Single-Phase Flow and Cluster Identification in Heterogeneous Pore Networks}

\author[inst1, inst2]{Ilan Ben-Noah}
\author[inst1]{Juan J. Hidalgo}
\author[inst1]{Marco Dentz}

\affiliation[inst1]{Institute of Environmental Assessment and Water Research (IDAEA), Spanish National Research Council (CSIC), Barcelona, Spain.}
\affiliation[inst2]{Department of Environmental Physics and Irrigation, Institute of Soil, Water and Environmental Sciences, The Volcani Institute, Agricultural Research Organization, Rishon LeZion, Israel}

\end{frontmatter}

\noindent\textbf{Contents of this file}
\begin{enumerate}
\item Figure SF1
\end{enumerate}

%
%



%

 \begin{figure}[b] 
 \includegraphics[width=\textwidth]{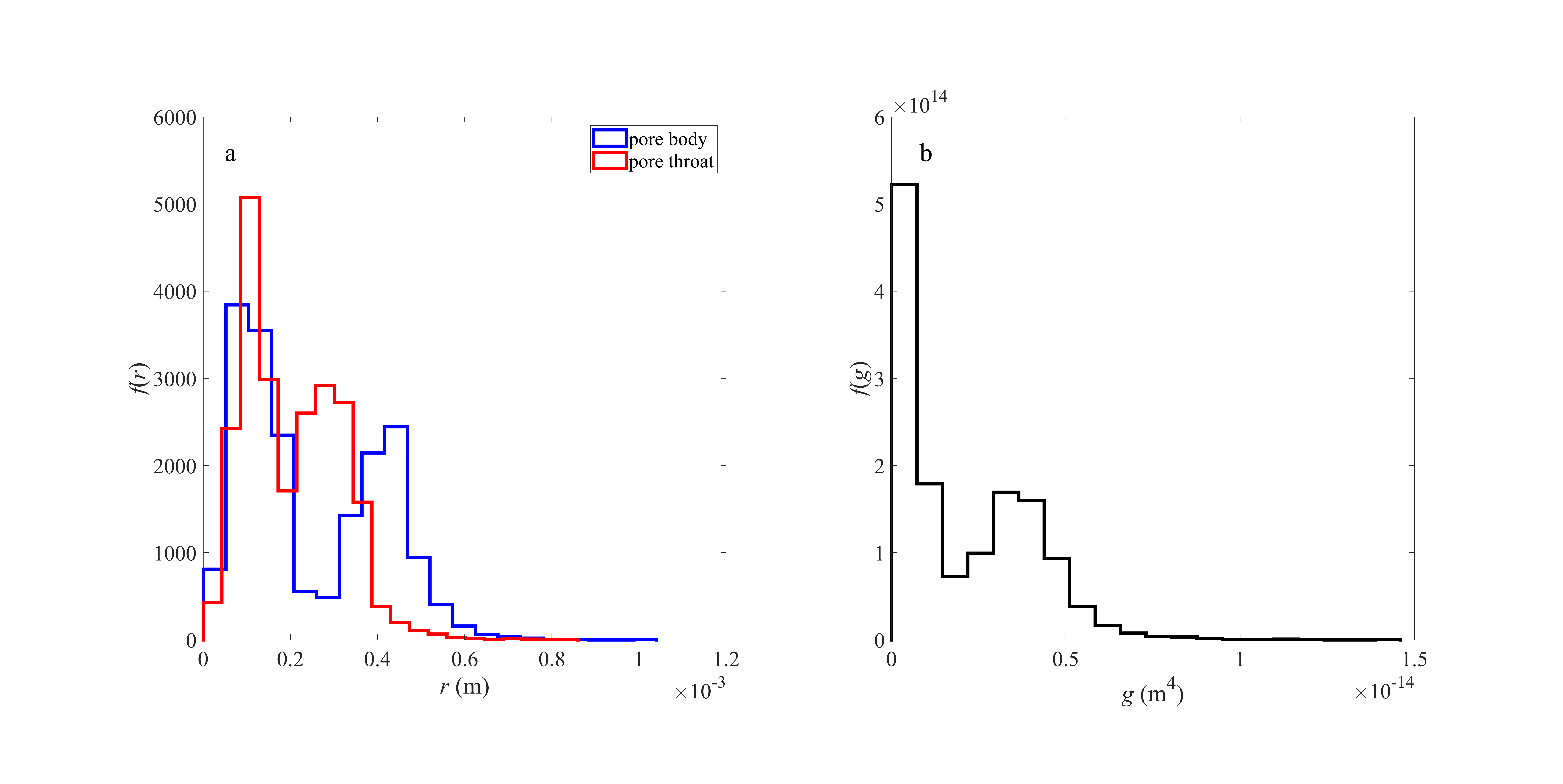}
 \caption{The probability density functions of a) the pore bodies and pore throats of the liquid phase from the non-processed image of the milifluidic device, and b) the bond conductivity, that is, $g=-Q_i\mu/\Delta P$, where $\Delta P$ is the pressure difference between the two sides of the bond, $Q_i$ the flow rate, and $\mu$ the dynamic viscosity. The milifluidic device is 105 mm long, 70 mm wide, and 0.5 mm thin. The solid medium comprises cylinders (pillars) with a mean diameter of $0.83$ mm (and standard deviation of $0.22$ mm), a height of $b=0.5$ mm, and a porosity of $0.72$ [14]. The pixel size of the 2D images is $\Delta x = 0.032$ mm. The pore space partitioning was evaluated following the method described in [16] using a Gaussian pyramid filter with a baseline standard deviation $\sigma_0=0.25$, an interval $\Delta \sigma = 0.005$, and a space scale factor of $\gamma=0.75$.}       
 \label{SF1}  
 \end{figure}